# Heterogeneous Catalysis on a Disordered Surface

L. Frachebourg, P. L. Krapivsky and S. Redner

*Center for Polymer Studies and Department of Physics*

*Boston University, Boston, MA 02215, USA*

(June 27, 1995)

We introduce a simple model of heterogeneous catalysis on a disordered surface which consists of two types of randomly distributed sites with different adsorption rates. Disorder can create a reactive steady state in situations where the same model on a homogeneous surface exhibits trivial kinetics with no steady state. A rich variety of kinetic behaviors occur for the adsorbate concentrations and catalytic reaction rate as a function of model parameters.

In modeling heterogeneous catalysis, it is often assumed that the catalytic surface is spatially homogeneous and that all surface sites have the same adsorption energy [1]. However, real surfaces exhibit some degree of spatial heterogeneity [2] which may be attributed to surface imperfections (dislocations, defects, ...), subsurface effects which modify interactions at the surface, or the arrangement of adsorbates on the surface and induced adsorbate-surface interactions.

In this Letter, we consider the effect of quenched surface imperfections on the reaction kinetics of the idealized monomer-monomer model [3]. Here $A$ and $B$ particles impinge upon a surface, with respective rates $k_A$ and $k_B$, and adsorb onto vacant sites. Nearest-neighbor adsorbed $AB$ pairs react and desorb with rate $k_r$, leaving behind two vacancies. When all surface sites have the same adsorption characteristics, this model has simple kinetics which can be obtained exactly by a mapping onto a kinetic Ising model with mixed zero-temperature Glauber and infinite-temperature Kawasaki dynamics [4]. For $k_A \neq k_B$, the adsorption imbalance leads to the surface quickly becoming saturated by the majority species and the reaction stops. For $k_A = k_B$, there is a fluctuation-induced coarsening of the surface into growing $A$ and $B$ adsorbed islands. The concentration of reactive interface, or nearest-neighbor adsorbed $AB$ pairs, $C_{AB}(t)$, vanishes with time as $t^{-1/2}$ for a one-dimensional "surface" [4] and as $(\ln t)^{-1}$ in two dimensions [5,6]. When islands become of the order of the size of the surface, there is no longer available $AB$ reactive interface and the reaction again stops. This saturation time is proportional to $L^2$ for an $L$-site chain and proportional to $L^2 \ln L$ for an $L \times L$ surface. Thus there is no reactive steady state on a homogeneous surface.

Our primary result is that surface disorder can create a reactive steady state. This state is different in character from that which arises in the monomer-dimer or ZGB model [7] for CO oxidation on metal surfaces. Further, a rich variety of kinetic behavior occurs which stems from the interplay between the surface disorder and the relative adsorption rates of the two species. Our model can be schematically represented by the following kinetic steps:

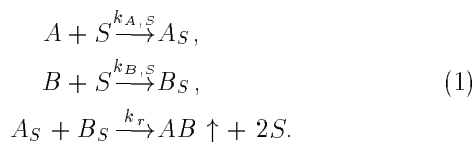

$$A + S \xrightarrow{k_{A,S}} A_S,$$
$$B + S \xrightarrow{k_{B,S}} B_S, \qquad (1)$$
$$A_S + B_S \xrightarrow{k_r} AB \uparrow + 2S.$$

The new feature is quenched surface disorder so that the adsorption rate of each species depends on the surface site, as well as on the incident particle fluxes (Fig. 1).

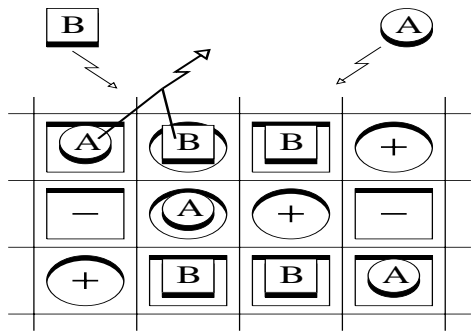

FIG. 1. Schematic illustration of the elemental processes on the disordered surface. Here + sites are represented by round indentations which preferentially adsorb (round) $A$ particles, while − sites (square indentations) preferentially adsorb (square) $B$ particles.

We choose dichotomous disorder for the surface sites: one type, +, with concentration $c_+$, prefers to adsorb $A$, while the complementary − sites, with concentration $c_- = 1 - c_+$, prefer to adsorb $B$. If the relative fluxes of $A$ and $B$ are $p$ and $q = 1 - p$, then the disorder is defined by the rule that a vacant + site becomes occupied by an $A$ with probability $p + \epsilon$ and by a $B$ with probability $q - \epsilon$, whereas a vacant − site becomes occupied by an $A$ with probability $p - \epsilon$ and by a $B$ with probability $q + \epsilon$. The strength of the adsorptive disorder is quantified by the parameter $\epsilon$ which obeys the constraint $0 \leq \epsilon \leq \min(p, q)$. For simplicity, we con-sider the reaction-limited process with adsorption rates much larger than the reaction rate ($k_{A,S}, k_{B,S} \gg k_r$). Thus whenever a nearest-neighbor $AB$ adsorbate pair reacts and desorbs, the two vacancies are immediately refilled. We have checked that similar behavior arises in the adsorption-limited process ($k_{A,S}, k_{B,S} \ll k_r$) and anticipate that our qualitative conclusions should still be valid for arbitrary reaction and adsorption rates.

To develop an understanding of the kinetics, consider first the mean-field limit. Define $x \equiv n_A^+/N$ and $y \equiv n_A^-/N$ as the respective concentrations of adsorbed $A$ on the $+$ and $-$ sites, where $N$ is the total number of surface sites. The respective concentrations of $B$ on the $+$ and $-$ sites are then $c_+ - x$ and $c_- - y$. The corresponding rate equations are,

$$\dot{x} = (c_+ - x)(x+y)(p+\epsilon) - x(1-x-y)(q-\epsilon),$$
$$\dot{y} = (c_- - y)(x+y)(p-\epsilon) - y(1-x-y)(q+\epsilon). \quad (2)$$

Here the reaction rate $k_r$ is absorbed into the time variable and overdot denotes the time derivative. In the first equation, for example, the first term accounts for the gain in $x$ due to a $B$ on a $+$ site (which is part of a nearest-neighbor $AB$ pair with the $A$ on either a $+$ or $-$ site) being replaced by an $A$.

For the homogeneous surface ($\epsilon = 0$), the surface quickly becomes filled by the majority species when $p \neq 1/2$, while for $p = 1/2$, the $A$ and $B$ concentrations remain static. For the disordered surface there are three steady state solutions: two trivial, $x = y = 0$ and $x = c_+$, $y = c_-$, corresponding to $B$- and $A$-saturation, and a nontrivial solution, $x = x_\infty(p, \epsilon, c_+)$, $y = y_\infty(p, \epsilon, c_+)$, corresponding to a reactive steady state [8]. Only one of these solutions is stable in the long-time limit as a function of the parameters $(p, \epsilon, c_+)$. Their relative stability is conveniently visualized within two-dimensional subspaces of the full $(p, \epsilon, c_+)$ phase space (Fig. 2(a)). In Fig. 2, the regions of $A$-saturated, $B$-saturated, and reactive states are denoted by **A**, **B**, and **R**, respectively. The phase boundaries are generally described by cumbersome expressions [8], but in the symmetric case $c_+ = 1/2$, the relatively simple formula holds

$$\epsilon = \sqrt{\frac{q(q-p)}{2}}, \quad \frac{1}{3} \leq p \leq \frac{1}{2}, \quad (3)$$

for the **B-R** boundary (with a symmetric expression for the **A-R** boundary). For the temporal behavior, it is easily found that the system approaches either saturation or a reactive steady state exponentially in time, except along the phase boundaries. However, along the **A-R** and **B-R** boundaries, the appropriate density asymptotically decays to zero as $t^{-1}$ [8].

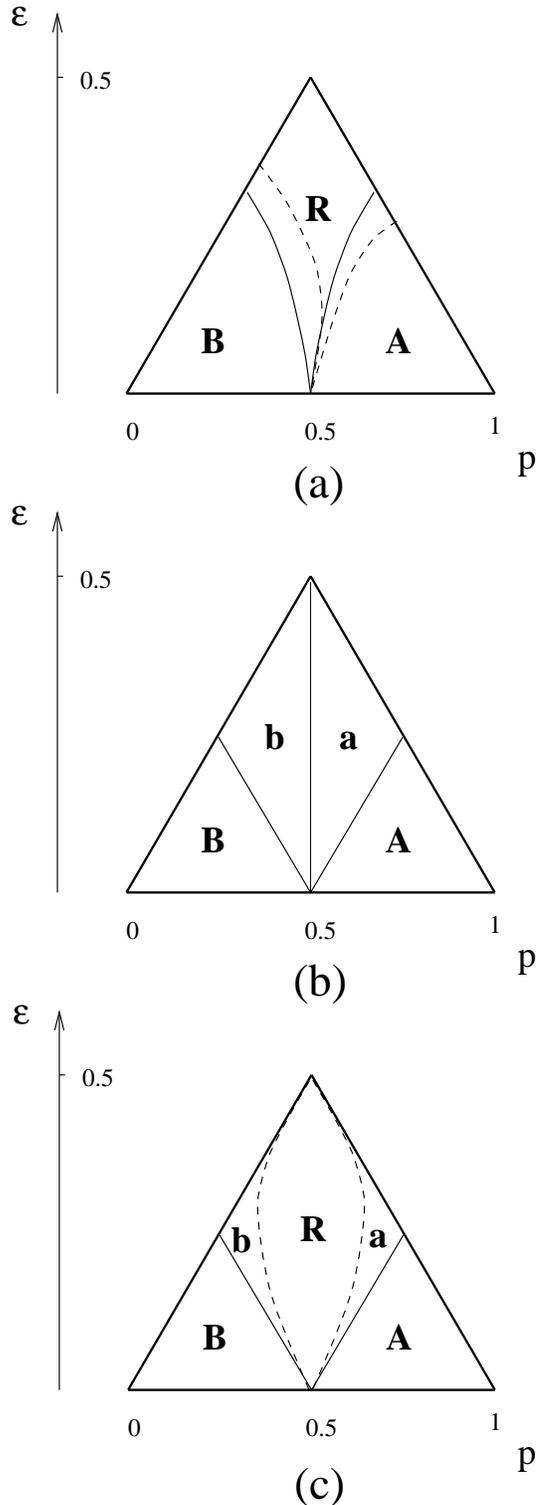

FIG. 2. Phase diagram of the disordered monomer-monomer model: (a) the mean-field approximation when $c_+ = \frac{1}{2}$ (solid lines) and $c_+ = \frac{1}{4}$ (dashed lines), (b) one dimension, and (c) two dimensions. The phases are indicated by bold letters. Exact phase boundaries are indicated by solid lines, while numerically determined phase boundaries are indicated by dashed lines.

For the disordered one-dimensional surface, several new regimes of behavior arise whose qualitative characters are independent of the values of $c_+$ and $c_-$ (Fig. 2(b)). For $p + \epsilon < 1/2$, $B$ is preferentially adsorbed on both the $+$ and $-$ sites and the concentration of adsorbed $A$, $C_A(t)$, decays exponentially in time. Simple bounds can be given for this decay. A lower bound arises by replacing the disordered system by a homogeneous one with $A$ adsorption probability equal to $p - \epsilon$ at all sites. Similarly, an upper bound arises by considering a homogeneous system with $A$ adsorption probability equal to $p + \epsilon < 1/2$. Since $C_A(t)$ relaxes exponentially to zero with time in these two models, we conclude that $p + \epsilon < 1/2$ indeed defines the **B**-phase of the disordered system with $C_A(t) \sim C_{AB}(t) \sim \exp(-at)$ with $a = a(p, \epsilon, c_+)$.

For $p = \epsilon = 1/4$, $-$ sites are permanently occupied by $B$, while the $A$ and $B$ adsorption probabilities both equal $1/2$ on the $+$ sites. The system therefore decouples into a succession of independent chains, each with fixed $B$ boundaries. The probability distribution for the length $L$ of these chains is $P(L) \sim \exp(-L/L_0)$ with $L_0 = -1/\ln(c_+)$. From the diffusive nature of the catalytic reaction [4], each of these chains will saturate with $B$ exponentially in time with a characteristic time of the order of $L^2$. Upon averaging over the distribution of chain lengths, the average concentration of $A$ is therefore given by

$$C_A(t) \sim \int_0^\infty \exp(-L/L_0) \exp(-Dt/L^2) \, dL,$$
$$\sim \exp\left(-\text{const.} \times (t/L_0^2)^{1/3}\right). \quad (4)$$

This decoupled chain problem is also equivalent to a one-dimensional random field Ising model with external field $h_i = \infty$ on a fraction $c_-$ of the sites and $h_i = 0$ on the remaining $c_+ = 1 - c_-$ sites. For Glauber dynamics, this problem has been solved by Forgacs et al. [9]. In principle, their approach could be extended to the mixed Glauber and Kawasaki dynamics of our catalysis model and would permit exact evaluation of the coefficient of the above stretched exponential.

On the **B-R** boundary $\epsilon = 1/2 - p$ (with $1/4 < p < 1/2$), the $+$ sites still have a probability of $1/2$ to adsorb both species, while the $-$ sites have a probability $0 < 2p - 1/2 < 1/2$ to adsorb $A$. Because of the global adsorption bias, the system still saturates to $B$. The lower bound $C_A(t) \propto \exp(-at^{1/3})$ can be immediately constructed by considering a system where $A$ does not adsorb on $-$ sites. While the construction of an upper bound is non-trivial, numerical simulations on a chain of $10^6$ sites show that the decay of $C_A(t)$ on this line still has the same form (4). Thus the **B-R** boundary belongs to the **B**-phase, although the asymptotic decay exhibits a stretched exponential behavior.

On the line $1/4 < \epsilon = p < 1/2$, the $-$ sites are still frozen with $B$, but there is a bias toward $A$ on the $+$ sites. This system is again equivalent to an ensemble of decoupled Ising chains (with mixed Glauber and Kawasaki dynamics), each with $-$ spins on the boundaries in a positive magnetic field. It is clear physically that the characteristic saturation time $\tau_L$ for a chain of length $L$ grows as $\exp(L)$. A Lifshitz argument of the type given in Eq. (4) then suggests a non-universal power law decay for $C_A(t)$. A rough estimate for the behavior of $\tau_L$ can be obtained by approximating the dynamics of a single chain with that of a biased random walk in the interval $0 < x < L$ with reflection at $x = 0$ and absorption at $x = L$, and with bias toward the reflecting barrier. The survival probability of this random walker is asymptotically given by $\exp(-t/\tau_L)$, with $\tau_L = (D/v^2) \exp(vL/D)$. To make contact with the disordered catalysis model, the bias should be given by $v = 4p - 1$ and the diffusion coefficient should be $D = 4p(1 - 2p)$. With this kinetics for a single chain, the Lifshitz argument for the average over all chains gives $C_A(t) \sim t^{-\alpha(p, c_+)} \times \ln t$, with $\alpha(p, c_+) = \ln(c_+) 2p(1 - 2p)/(1 - 4p)$. This naive approximation agrees well with our numerical simulations [8].

For $1/4 < p < 1/2$, $1/2 - p < \epsilon < p$ (region **b** in the phase diagram), the average adsorption rate globally favors $B$ but $+$ sites adsorb $A$ with a probability greater than $1/2$. This dichotomy again leads to a power law decay of $C_A(t) \sim t^{-\beta}$, but with a non-universal exponent $\beta(p, \epsilon, c_+)$. By comparison to the results for the case $p = \epsilon$, where the $-$ sites are permanently occupied by $B$, the bound $\beta(p, \epsilon, c_+) \leq \alpha((p + \epsilon)/2, c_+)$ can be established. We have numerically determined the exponent $\beta$ for various parameter values, and a qualitatively similar parameter dependence as the exponent $\alpha$ is found. We define this region of the phase diagram as the **b**-phase to indicate that it still corresponds to $B$ saturation but with power-law rather than exponential kinetics (as in the **B** phase).

Finally, on the line $p = 1/2$, the $A$ and $B$ concentrations are marginally stable and a saturated state is reached by the formation of a coarsening mosaic of approximately equal size $A$ and $B$ domains. (Here we assume $c_+ = c_- = 1/2$, otherwise the **a-b** boundary will be curved.) However, the disorder drastically slows down the interface evolution compared to the homogeneous system. To appreciate the effect, consider the motion of a single $AB$ interface. In a region with contiguous $+$ sites, the interface experiences a bias to the right, while in a contiguous region of $-$ sites, the interface is biased to the left. Thus there are effective potential wells for $AB$ interface motion with minima at points where a $+-$ nearest-neighbor pair of sites occurs. Thus the interface motion is analogous to Sinai diffusion [10]. and the density of interfaces should decay as However, there are microscopic differences in details between Sinai diffusion and our disordered catalysis model, and the interface density is numerically found to decay as

$$C_{AB}(t) \sim (\ln t)^{-\gamma(\epsilon)}, \qquad (5)$$

with a non-universal exponent $\gamma(\epsilon)$, in contrast to $\gamma = 2$, if the analogy with Sinai diffusion was exact.

For a two-dimensional disordered surface, five phases occur, including a reactive steady state **R**. As in one dimension, it may established rigorously that in the **A** phase, where $A$ is preferentially adsorbed on both the $+$ and $-$ sites, $A$-saturation is approached exponentially in time. This conclusion is independent of the values of $c_+$ and $c_-$. In numerical simulations on a square lattice of $10^6$ sites, we have also found **a** and **b** phases for the symmetric case $c_+ = c_- = 1/2$ (Fig. 2(c)). In these regions the minority species decays as a power law in time, $C_{\text{minority}}(t) \sim t^{-\beta_m}$, with a nonuniversal exponent $\beta_m = \beta_m(p, \epsilon, c_+)$. Finally, in the reactive steady state (**R**) the densities approach nonzero values exponentially in time.

More subtle kinetics arises on the phase boundaries. On the **a-A** and **b-B** boundaries the minority species appears to exhibit a stretched exponential decay, $\ln(C_{\text{minority}}(t)) \sim -t^{\delta_m}$. To obtain a theoretically plausible estimate for this exponent, it is convenient to consider the extremal point $p = \epsilon = 1/4$ (or $q = \epsilon = 1/4$) and follow the same line of reasoning as in one dimension. For $p = \epsilon = 1/4$, for example, the $-$ sites are permanently occupied by $B$, while on the $+$ sites, the $A$ and $B$ adsorption probabilities both equal $1/2$. The $-$ sites thus play the role of "sources" of the $B$-phase and the long-time behavior is determined by the filling of large voids which do not contain $-$ sites. Within the voids the kinetics is just that of the homogeneous and unbiased monomer-monomer model (i.e., $p_{\text{eff}} \equiv p + \epsilon = 1/2$, $q_{\text{eff}} \equiv q - \epsilon = 1/2$). By again appealing to the mapping onto onto a kinetic spin model [5], it may be argued that the average concentration of $A$ in a void decreases as $\exp(-t/R^2)$. Since the probability distribution for disk-shaped voids of radii $R$ is Poissonian, $P(R) \sim \exp(-R^2/R_0^2)$, the configuration average $A$ concentration is

$$C_A(t) \sim \int_0^\infty \exp\left(-\frac{R^2}{R_0^2} - \frac{t}{R^2}\right) dR \sim \exp\left(-t^{1/2}\right). \qquad (6)$$

Therefore $\delta_m = 1/2$ at the extremal points. Based on the equivalence between catalysis kinetics and the trapping problem in two dimensions, we expect the universal value $\delta_m = 1/2$ along the **a-A** and **b-B** boundaries.

Finally, on the **a-R** and **b-R** boundaries we observe very slow decay of the minority species. It is difficult to establish whether this decay is a power-law with tiny exponents or logarithmic. The logarithmic behavior seems more natural and qualitatively agrees with the $(\ln t)^{-1}$ decay for the number of interfaces which is expected to occur at the extremal point $p = 1/2$, $\epsilon = 0$ (see [5]).

In summary, we have found a reactive steady state in the monomer-monomer model of heterogeneous catalysis on surfaces with adsorptive disorder. This result supports intuitively appealing hypothesis that disorder generally stabilizes the reactive steady state. We have also observed rich kinetics for the reactant concentrations and reactivity in our catalysis model, including exponential, stretched exponential, power law, and logarithmic decays with time for different system parameters.

We gratefully acknowledge the financial support of the Swiss National Foundation (to LF), and ARO grant DAAH04-93-G-0021 and NSF grant DMR-9219845 (to PLK and SR).